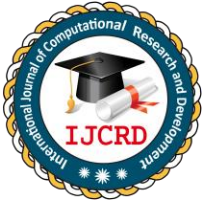

# EFFECT OF IMPLEMENTATION OF IMPROVED METHODS OF THE LIFE CYCLE STAGES ORGANISATION TO THE ONLINE COMMUNITY MANAGEMENT

**Yuriy Syerov*, Olha Trach** & Solomia Fedushko***

\* Philosophy Doctor of Technical Sciences Associated Professor, Deputy Head of Research Work of Social Communications and Information Activities Department, Lviv Polytechnic National University, Lviv, Ukraine

\*\* Assistant Professor, PG Scholar, Social Communications and Information Activities Department, Lviv Polytechnic National University, Lviv, Ukraine

\*\*\* Philosophy Doctor of Technical Sciences, Assistant Professor, Social Communications and Information Activities Department, Lviv Polytechnic National University, Lviv, Ukraine

**Abstract:**
This paper presents the current problem of investigation of the effect of implementation of improved methods of the life cycle stages organisation to online community management. The online community life cycle is the sum of the stages of organisation and development of online community. The current approaches of scientific researches of social processes within the WWW are analysed. The types of life cycles are distinguished and implemented in the online community management. The algorithm of life cycle stages of the online community is designed. The total quality of the life cycle stages execution and quality of the life cycle stages execution of Lviv Polytechnic online community from 2013 till 2016. Currently, the appropriate study and development of community management hardware and software provoke the most interest because online communities are a widespread and popular phenomenon, and the existing management software for them is imperfect and not complex.

**Index Terms:** Online Community, Life Cycle Stages, WWW, Management & Efficiency

## 1. Introduction:

At the current period of improvement and modernisation of online community management and Internet and global information system World Wide Web, in general, online communities consisting of usual users [1-3] of network are gaining an increasing role. Online community [4-6] is a social group of web users who communicate and co-operate via the Internet with the help of specialized services and websites within the WWW. Online communities are one of the basics for the up-and-coming establishment of the modern information society. The United Nations recognized information society as the priority target of modern human development. The information society is defined as a society, where everyone is able to create information and knowledge, have access to it, use and share it, giving the ability to individuals, communities and peoples to fully realize their potential, to promote their sustainable development and to improve their quality of life.

In scientific researches of social processes within the WWW one can distinguish the following approaches:

**Satisfaction of Consumer's Demands** – this approach comprises the research of website users' behavior (web usage mining) and issues of website design.The first approach combines researches [7-9 ] on web usage mining (works of R. Kosala, R. Bayeza-Yates, G. Piatetsky-Shapiro), user activity monitoring (papers of B. Mobasher, S. Chakrabarti), website usability (J. Nielsen, S. Krug).

**Communication Systems** – this approach includes the research of various communication systems of WWW users: e-mail, instant messenger, audio/video communication tools etc. The second approach includes organization and development of hardware and software tools for exchanging information among people on the Internet. Basic concepts of the development of this approach [10-13] were defined by T. Berners-Lee and J. Hendler.

**Formation of Communities and their Management** – this approach investigates the social aspects of human behavior and the development of community management hardware and software. The third approach consists of two parts – investigation of socio-psychological aspects of human communication – social engineering (there are known papers on this field by following scientists [14-17] – K. Popper, A. Podgorecki, J. Karpinski) and the organization of online community management hardware and software.

Currently, the appropriate study and development of community management hardware and software provoke the most interest because online communities are a widespread and popular phenomenon, and the existing management software for them is imperfect and not complex.

As a result, among the known types of online communities, web forums are one of the most powerful and popular services of WWW, which are designed for arranging users' communication. The web forums are a unique source of information, a place of accumulation of large amounts of important, valuable information and knowledge, a stimulus for various commercial and public projects.

**2. Implementation of Improved Methods of the Life Cycle Stages Organisation of the Online Community Management:**

The online community life cycle is the sum of the stages of organisation and development of online community.

Considering the existing information about the online community should pay attention to the following types of life cycles, namely:
- ✓ Software - online community as software. The life cycle of the software describes in the form of methodologies or group standards [18-22];
- ✓ Web site - an online community as an information system [23];
- ✓ Investment project - features of commercial or investment project. The researchers of the project lifecycle investment project lean on the program of industrial development of UNIDO and project management methodologies [24];
- ✓ Basis management and marketing of products - marketing and management needs in the management of the virtual community [25].

Based on the existing compatible life cycles to the needs and models online communities can highlight the following stages of the life cycle of online communities (see Figure 1):

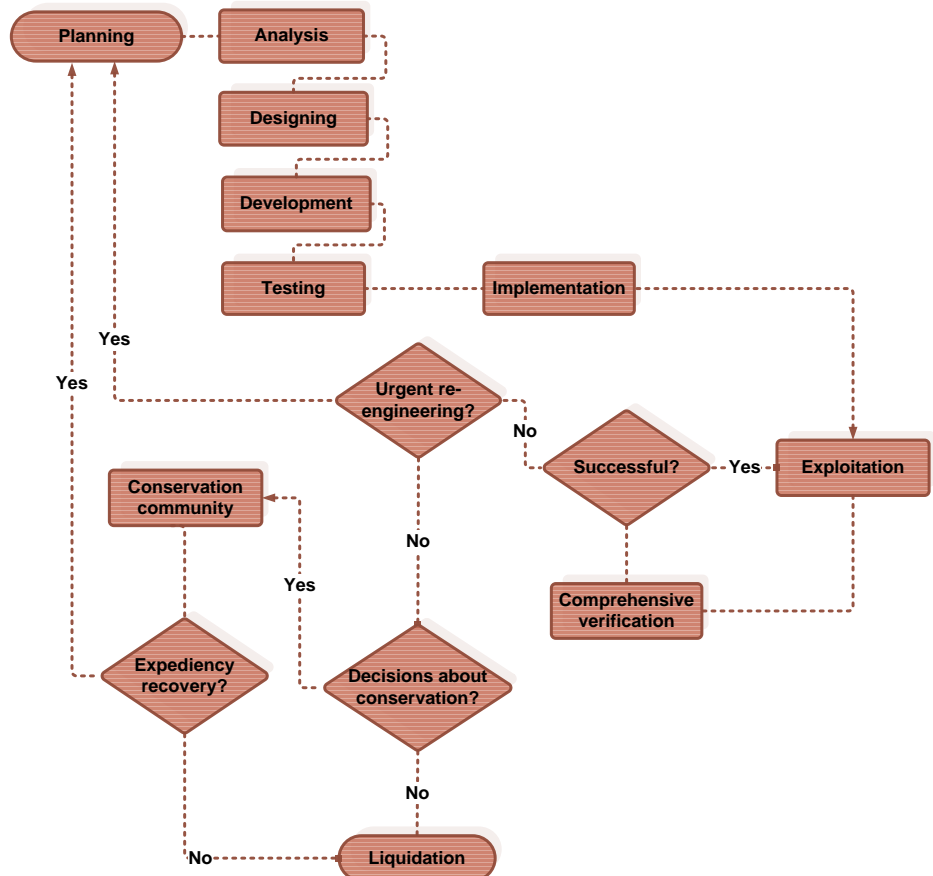

Figure 1: The Algorithm of Life Cycle Stages of the Online Community

The algorithm of life cycle stages of the certain online community consists in carrying out of such acts as:

**Planning** – stage of planning the online community development, prior formulation of the problem to the creating, launch and management of online community.

**Analysis** – analysis of existing online communities (including content and users) and competitiveness with them, quality decision making concerning expediency of a virtual community.

**Designing** – online communities designing based on the task and specifications, analysis, creation of conditions and types of online communities.

**Development** – development of online communities based on the preliminary design.

**Testing** – online communities testing for finding and correcting errors, online communities testing involve the division into internal and public testing (Alpha and the Beta testing).

**Implementation** – starting online communities is considered as the introduction it in the Internet and increasing the number of community members.

**Exploitation** – development, support and management of the online community.

**Comprehensive Verification** - external process, during this process specification of compliance with all the set goals and objectives of its already established and operating this online community is carried out.
**Successful Checking** – the successful comprehensive verification checking. With the successful operation of verification community returns to step exploitation and continues to develop. Upon detection a large competitiveness or other errors we must make a decision on urgent re-engineering.
**Urgent Re-Engineering** – a return to the stage of analyzing otherwise the while making a positive decision on urgent re-engineering, arises necessity of a decision on conservation of the online community.
**Decisions About Conservation** – decisions on conservation of the online community. If the decision is negative, then need came in to complete elimination of the virtual community. If a positive decision is necessary return to the stage of analysis.
**Conservation Community** – conservation of online community.
**Expediency Recovery** – online community is dormant until it is necessary to decision making about expedience of recovery.
**Liquidation** – the complete elimination of the online community.

Taking into consideration highlighted stages of algorithm of life cycle stages of the certain online community, the life cycle of *i*-th online community is defined, it is as follows:

$$LifeCycle_{Com_i} = \left\langle \begin{array}{l} Plan_{Com_i}, Analys_{Com_i}, Design_{Com_i}, Devel_{Com_i}, Test_{Com_i}, \\ Implement_{Com_i}, Expl_{Com_i}, ComVer_{Com_i}, ExpRec_{Com_i}, Liq_{Com_i} \end{array} \right\rangle \quad (1)$$

Where $Plan_{Com_i}$ – planning the online community development ;

$Analys_{Com_i}$ –analysis of existing expediency recovery;

$Design_{Com_i}$ –online communities designing;

$Devel_{Com_i}$ –development of online communities;

$Test_{Com_i}$ –online communities testing for finding and correcting errors;

$Implement_{Com_i}$ –online communities implementation;

$Expl_{Com_i}$ –online communities exploitation;

$ComVer_{Com_i}$ –comprehensive verification;

$ExpRec_{Com_i}$ –expediency recovery of online community;

$Liq_{Com_i}$ – liquidation of online community.

The total quality of the life cycle stages execution of Lviv Polytechnic online community is calculated by using the formula (1). The result of calculating total quality of the life cycle stages execution of Lviv Polytechnic online community is shown in Figure 2.

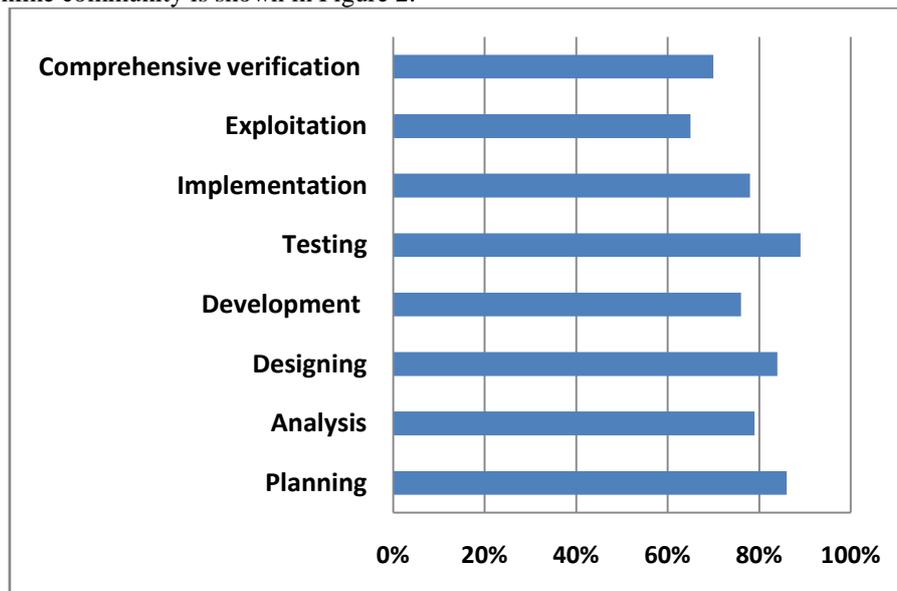

Figure 2: The Total Quality of the Life Cycle Stages Execution of Lviv Polytechnic Online Community

The quality of the life cycle stages execution of Lviv Polytechnic online community from 2013 till 2016 shows in Figure 3.

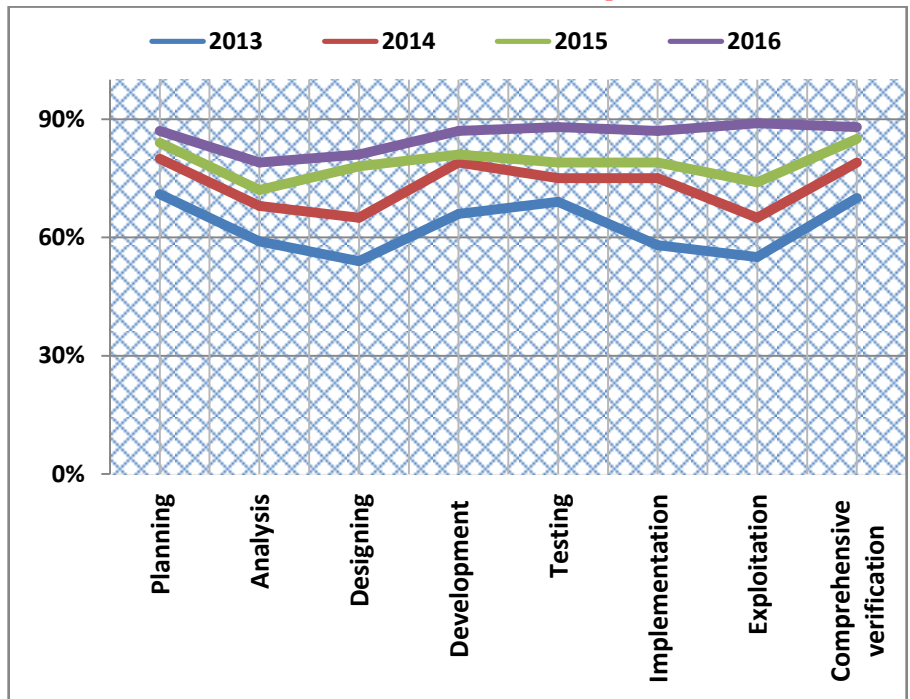

Figure 3: The Quality of the Life Cycle Stages Execution of Lviv Polytechnic Online Community from 2013 till 2016

**3. Conclusion:**

The paper decided an important scientific task of developing methods and tools for implementation of improved methods of the life cycle stages organisation to online community management using author's developed theoretical foundations and software system that provides effective functioning of the community in the long term. At present the growth of number and measurement of online communities and the incessant development of the existing ones are monitored. Significant amounts of users and data, organizations necessity of management for satisfaction of business requirements, advertising and commercial benefits the projected process management of the online community is required. The existing studies of organization of the life cycle of online communities are underdeveloped and incomplete.This fact causes a necessity of development of mathematical and computer software of the life cycle of online communities organization and it will be possible to fix of already existent disadvantages of online communities, the possibility to predict the development of new online communities, to improve controllability and level of ensuring the needs of founders and customers of the online community.